\documentclass[aps,prb,superscriptaddress,twocolumn]{revtex4-1}
\usepackage{graphicx}
\usepackage{amssymb}
\usepackage{amsmath}
\usepackage{multirow}
\usepackage{float}
\graphicspath{{images/}}
\usepackage[usenames, dvipsnames]{color}

\begin{document}

\title{Multi-component order parameter superconductivity of Sr$_2$RuO$_4$ revealed by topological junctions}

\author{M. S. Anwar} \altaffiliation[Current address:~]{London Centre for Nanotechnology, University College London, UK.}
\email[E-mail:]{s.anwar@ucl.ac.uk}\affiliation{Department of Physics, Kyoto University, Kyoto 606-8502, Japan}

\author{R. Ishiguro} \affiliation{Department of Mathematical and Physical Sciences, Faculty of Science, Japan Women's University, Tokyo 112-8681, Japan} \affiliation{Department of Applied Physics, Faculty of Science, Tokyo University of Science, Katsushika  Tokyo 162-8601, Japan}

\author{T. Nakamura} \affiliation{Department of Physics, Kyoto University, Kyoto 606-8502, Japan}\affiliation{Institute for Solid State Physics, the University of Tokyo, Kashiwa 277-8581, Japan}

\author{M. Yakabe}\affiliation{Department of Applied Physics, Faculty of Science, Tokyo University of Science, Katsushika  Tokyo 162-8601, Japan}

\author{S. Yonezawa}\affiliation{Department of Physics, Kyoto University, Kyoto 606-8502, Japan}

\author{H. Takayanagi}\affiliation{Department of Applied Physics, Faculty of Science, Tokyo University of Science, Katsushika  Tokyo 162-8601, Japan}

\author{Y. Maeno}\affiliation{Department of Physics, Kyoto University, Kyoto 606-8502, Japan}

\date{\today}

\begin{abstract}

Single crystals of the Sr$_2$RuO$_4$-Ru eutectic system are known to exhibit enhanced superconductivity at 3~K, in addition to the bulk superconductivity of Sr$_2$RuO$_4$ at 1.5~K. The 1.5-K phase is believed to be a spin-triplet, chiral $p$-wave state with the multi-component order parameter, giving rise to chiral domain structure. In contrast, the 3-K phase is attributable to enhanced superconductivity of Sr$_2$RuO$_4$ in the strained interface region between Ru inclusion of a few to tens of micrometers in size and the surrounding Sr$_2$RuO$_4$. We investigate the dynamic behavior of a topological junction, where a superconductor is surrounded by another superconductor. Specifically, we fabricated Nb/Ru/Sr$_2$RuO$_4$ topological superconducting junctions, in which the difference in phase winding between the $s$-wave superconductivity in Ru micro-islands induced from Nb and the superconductivity of Sr$_2$RuO$_4$ mainly governs the junction behavior. Comparative results of the asymmetry, hysteresis and noise in junctions with different sizes, shapes, and configurations of Ru inclusions are explained by the chiral domain-wall motion in these topological junctions. Furthermore, a striking difference between the 1.5-K and 3-K phases is clearly revealed: the large noise in the 1.5-K phase sharply disappears in the 3-K phase. These results confirm the multi-component order-parameter superconductivity of the bulk Sr$_2$RuO$_4$, consistent with the chiral $p$-wave state, and the proposed non-chiral single-component superconductivity of the 3-K phase.
\end{abstract}

\maketitle

\section{Introduction}

Spin-triplet superconductivity is rich in physics due to its spin and orbital degrees of freedom compared to ordinary spin-singlet superconductivity. The layered perovskite oxide Sr$_2$RuO$_4$ (SRO) is one of the leading candidates of spin-triplet superconductors (TSCs) with superconducting transition temperature ($T_{\textrm {c}}$) of 1.5~K~\cite{Maeno2012}. Since the discovery of superconductivity in SRO~\cite{Maeno1994}, an intensive amount of experimental and theoretical works have been performed to understand the nature of its superconducting order parameter~\cite{Machenzie2003, Maeno2012,Anwar2015,Anwar2016}. 

Superconductivity of SRO is extremely sensitive even to non-magnetic impurities~\cite{Nikugawa2003}. Spin susceptibility measurements by the nuclear magnetic resonance (NMR)~\cite{Ishida1998} and the polarized neutron scattering~\cite{Duffy2000} below $T_{\textrm{c}}$ supports the spin-triplet scenarios. Recently, invariant spin-susceptibility is re-confirmed using Ru and Sr nuclei and O as well~\cite{Ishida2015,Manago2016}. The muon spin rotation and magneto-optical Kerr effect evidence the time-reversal symmetry breaking in the superconducting order parameter of SRO~\cite{Xia2006}. These observations support the chiral $p$-wave spin-triplet nature of the order parameter that can be represented as $d=\hat{z}(k_x \pm ik_y)$. Recent observations of long range proximity effect emerged at SrRuO$_3$/SRO interface also supports the spin-triplet scenario for SRO~\cite{Anwar2015,Anwar2016}. Because of the orbital phase winding, it is believed that SRO is a typical example of topological superconductors, and gapless states consisting of Majorana fermions are expected to emerge at its boundaries~\cite{Matsumoto1999,Read2000,Stone2004,Qi2011,Kashiwaya2011,Alicea2012,Tanaka2012}. Extensive theoretical work~\cite{Maeno2012} also supports that the nature of the orderparameter of SRO is chiral $p$-wave spin-triplet with broken time reversal symmetry, although there are still unresolved issues~\cite{Hicks2010,Yonezawa2013,Hassinger2017}.
\begin{figure*}
	\begin{center}
		\includegraphics[width=12cm]{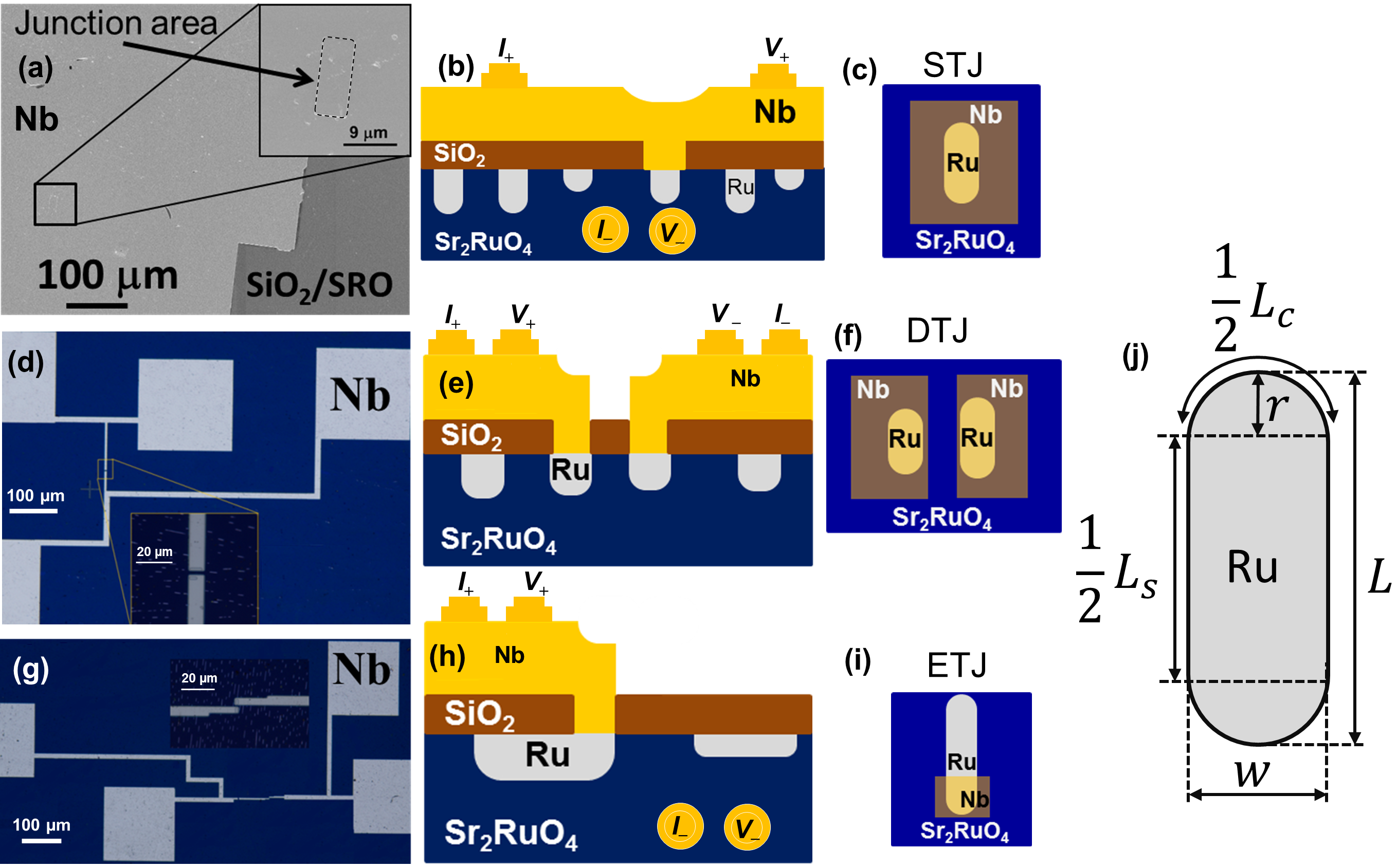}
		\caption{Three kinds of topological junctions investigated in this study. (a) Scanning electron microscope (SEM) image of a single topological junction (STJ) fabricated using a single Ru-inclusion. The inset is a magnified image showing a window over Ru-inclusion. Schematic cross-sectional image (b) and top-view (c) of the STJ. (d) Optical microscope image of a double topological junction (DTJ) prepared by depositing Nb electrodes onto two parallel Ru inclusions. The inset shows the junction area. Schematic side (e) and top (f) views of the DTJ. (g) Optical microscope image of an edge topological junction (ETJ) made by depositing Nb electrode only at the edge part of a Ru inclusion. A magnified image around the junction area is shown in the inset. Schematic images of side (h) and top (i) views of the ETJ. (j)Schematic illustration of a single elliptical Ru-inclusion of length $L$ and width $w$. The total length of straight and curved parts are $L_s$ and $L_c$, respectively.}
		\label{devices}
	\end{center}
\end{figure*}

Two-fold degeneracy of chirality leads to the formation of chiral domains with clockwise $(k_x + ik_y)$ and anticlockwise $(k_x - ik_y)$ chirality in the orbital order parameter. As a result, two chiral domains are naturally separated by a chiral domain wall (chiral-DW), like ferromagnetic domain walls in a ferromagnetic material. Recently, the existence of chiral-DWs and its dynamic behavior has been investigated experimentally and theoretically~\cite{Kindwingira2006,Xia2006,Kambara2008,Kambara2010,Hicks2010,Anwar2013,Saitoh2015,Bouhon2010,Nago2016}. Kindwingra $et$ $al.$~\cite{Kindwingira2006}, reported complicated and hysteretic diffraction patterns in Pb/Cu/SRO junctions. They explained their data in the scenario of chiral-DW dynamics and suggested the chiral-DW size of the order of 1~$\mu$m. The size of a chiral domain is controversial and it may depend on the experimental probe, e.g. experiments using Kerr effect~\cite{Xia2006}, probing the bulk time-reversal symmetry breaking, and scanning SQUID~\cite{Hicks2010}, probing a local magnetic field due to the edge current, suggested 50~$\mu$m and 0.4~$\mu$m, respectively. It is difficult to estimate the upper limit that can also depend on quality of the sample. However, recently Saitoh $et$ $al.$~\cite{Saitoh2015}, studied the ``inversion symmetry'' in the magnetic field dependence of the critical current of Nb/SRO junctions while varying the junction size, and estimated the domain size to be 5~$\mu$m. The value is similar to the value proposed in our previous report~\cite{Anwar2013}, in which we investigated Nb/Ru/SRO topological junctions and observed the telegraphic noise attributable to chiral-DW motion~\cite{Anwar2013}. 

A ``topological junction'' consists of a superconductor that is surrounded by another superconductor in such a way that the difference in phase winding mainly dictates the junction behavior~\cite{Anwar2013,Nakamura2011,Nakamura2012}. Intuitively, transport properties of a topological junction where a spin-singlet superconductor (SSC) is surrounded by a chiral $p$-wave TSC can also detect the dynamic behavior of the order parameter of TSC. We have already reported such dynamic behavior in topological junctions that are fabricated using naturally existing Ru metal inclusions inside SRO-Ru eutectic single crystals. Spin-singlet superconductivity is induced into Ru by putting Nb ($T_{\textrm{c}}$=9~K) directly onto a Ru inclusion. For utilization of chiral-DWs of SRO based junctions, it is now important to investigate the controllability of chiral-DW motion. In this article, we report our systematic study of current voltage ($I$-$V$) characteristics of Nb/Ru/SRO junctions fabricated in different configurations. It is observed that chiral domain dynamics is strictly related to the geometry and size of junctions.

\section{Experimentation}

We fabricated Nb/Ru/SRO micron-sized superconducting topological junctions using SRO-Ru eutectic crystals. Typically the width of a Ru inclusion is $\approx$ 2~$\mu$m, the length is of the order of 1-50~$\mu$m, and the depth is about 10~$\mu$m~\cite{Maeno1998}. The crystals were grown using a floating zone method~\cite{Mao2000}. In a SRO-Ru eutectic crystal, the onset $T_{\textrm{c}}$ is significantly enhanced up to 3~K. This superconducting phase with enhanced $T_{\textrm{c}}$ is known as the 3-K phase~\cite{Maeno1998,Yaguchi2003}. As suggested by various measurements~\cite{Yaguchi2003,Maeno2012}, the 3-K phase emerges at the SRO/Ru interface, possibly because of induced local strain on SRO side as demonstrated using pure SRO under uniaxial strains~\cite{Hicks2014,Taniguchi2015,Hicks2017}. 

Rectangular SRO-Ru substrates of the size 3$\times$3$\times$0.5~mm$^3$ were prepared by cutting the crystal along the $ac$-plane and cleaving it along the $ab$-plane. Note that the $ab$-surface of SRO does not show a good electrical contact with metals like Nb, Pb, etc., because of its bad adhesion and possible surface reconstruction of the Ru-O octahedra \cite{Matzdorf2000}. However, Ru metal inclusions provide a good adhesion to develop a good electrical contact. On the other hand, epitaxial growth of a materials with relatively lower crystal mismatch on SRO can also improve the electrical contact~\cite{Anwar2015,Anwar2016}. More importantly, a Ru inclusion naturally provides an embedded metal surrounded by a TSC, being quite suitable to develop a topological junction. For these reasons, we used SRO-Ru eutectic crystals.

\begin{figure}
	\begin{center}
		\includegraphics[width=7.4cm]{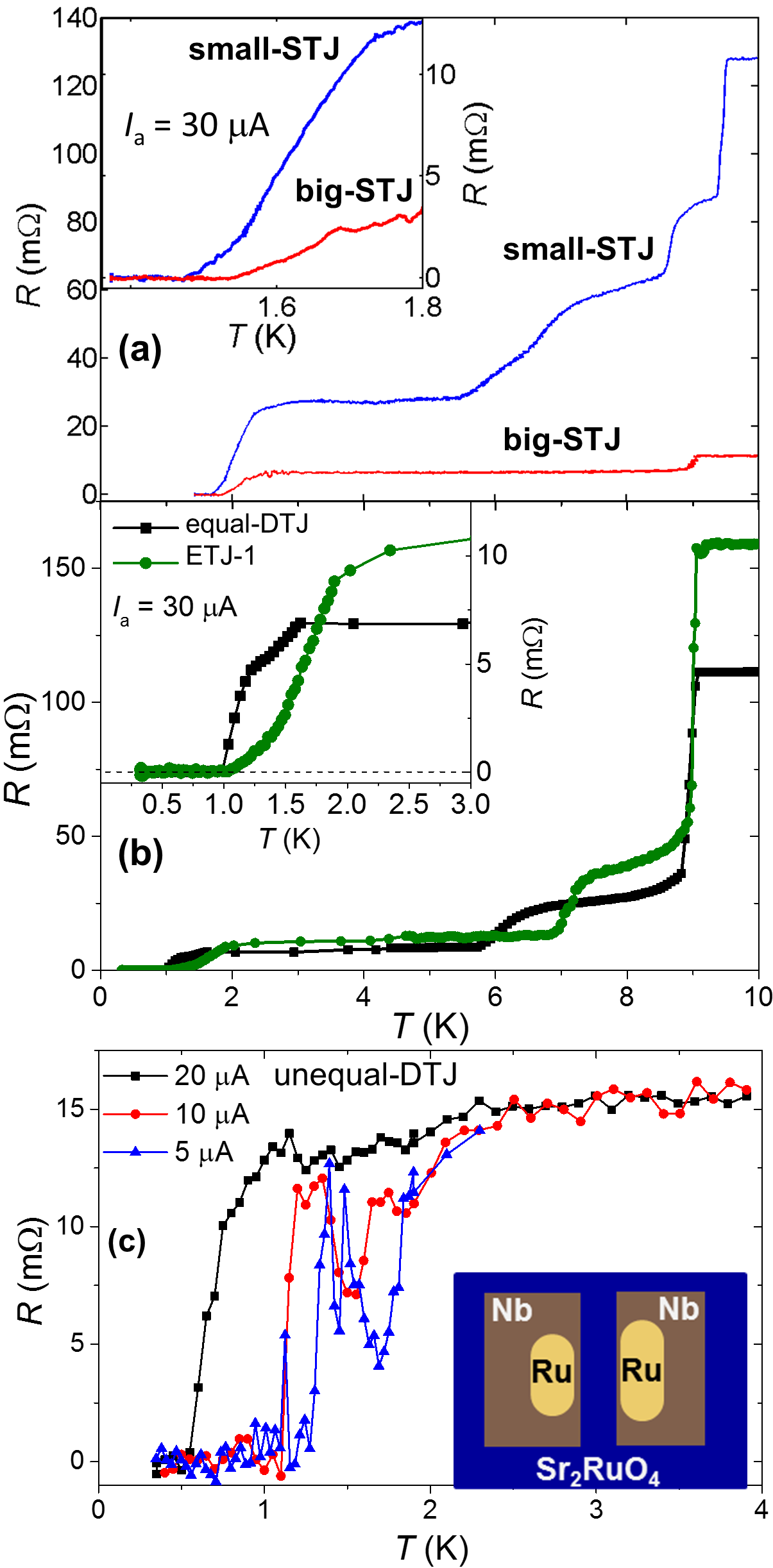}
		\caption{Junction resistance as a function of temperature. (a) Resistance vs temperature of STJs. Superconducting transitions corresponding to Nb and SRO are observed. (b) Temperature dependent resistance of the equal-DTJ (black points) and ETJ (green points). Data are collected using an Oxford ${}^3$He cryostat. Inset shows the resistance at temperatures close to the transition to zero resistance. (c) Resistance vs temperature of the unequal-DTJ measured at different applied current. Inset shows the schematic top view of the junction. These data are obtained in using a PPMS ${}^3$He probe.}\label{RT}
	\end{center}
\end{figure}

Next, the $ab$-surface of SRO substrates was polished using diamond slurry of 0.25-$\mu$m average particle size. Shortly after, a 300-nm thick SiO$_x$ layer was deposited on the $ab$-surface using RF sputtering technique with a backing pressure of $\approx$ 10$^{-7}$~mbar. Such a thick insulating layer guarantees the prevention any pin-holes to provide a short between Nb and Sr$_2$RuO$_4$. Then we coated the substrate with the photoresist (TSMR-8800) and exposed the photoresist with maskless lithography based on UV-laser only selected areas to achieve targeted junction geometries. The exposed resist was removed with TMAH2-83$\%$ developer for 120~sec. The substrate was rinsed with DI-water for 30~sec and dried with N$_2$ gas. The exposed part of the SiO$_x$ layer was etched away with CHF$_3$ gas, which opened a window over a desired area that we want to use as a junction area. In this process, a fluoride thin film may be generated on the surface of the sample. Therefore, we performed O$_2$ plasma cleaning of the junction area. The resist was removed using N-methyl-2-pyrolidone (NMP) and the substrate was cleaned with acetone and isopropanol. In the next step, we prepared the Nb electrodes using a lift-off technique with bilayer photoresist (LOR-10A and TSMR-8800) and by UV-laser photolithography. Ar-ion etching was performed {\it in situ} immediately prior to deposition of Nb layer in order to remove newly formed RuO$_x$ layer. A 600-nm thick Nb layer was sputtered with a base pressure of $10^{-9}$~mbar. Finally, the lift-off was accomplished with NMP.

It is a natural question whether we can control the dynamic behavior by changing the junctions configurations. To find the answer, we investigated the topological junctions with different configurations. The first type is single topological junctions (STJs), where a Nb electrode was deposited over a full single Ru-inclusion as presented in Fig.~\ref{devices}(a-c). For these junctions, two electrical leads ($\phi$ 25~$\mu$m gold wires) were connected with Nb electrode and other two leads were connected directly to a side of the eutectic crystal as shown in Fig.~\ref{devices}b. We prepared two STJs, one with a 20-$\mu$m long Ru inclusion (big-STJ) and another with a 5-$\mu$m long Ru inclusion (small-STJ). The second type is double topological junctions (DTJs), in which two separate Nb electrodes were deposited on two parallel and 4-$\mu$m-apart Ru inclusions (see Fig.~\ref{devices}(d-f)). For DTJs, two electrical leads were connected to each of the Nb electrodes. We prepared two different DTJs: equal-DTJ, where both Ru-inclusions are of the same size ($\approx$ 6~$\mu$m long) and unequal-DTJ, where one Ru-inclusion is 4-$\mu$m long and the other is 6-$\mu$m long. The third type is edge topological junctions (ETJs), in which one Nb electrode was deposited only on the edge of Ru inclusion ($3\times2~\mu$m$^2$), as shown in Fig.~\ref{devices}(g-i). The lead configuration of  this type of junction is similar to that for the STJ (Fig.~\ref{devices}(h)). Figure~\ref{devices}j shows schematically a rectangular with round tips Ru-inclusion with total length $L$ and width $w$. The total length of the straight part is \(L_{\textrm{s}}=2L_{\textrm{s}}=2(L-2r)=2(L-w)\) and total length of curved part is \(L_{\textrm{c}}=2\pi r=\pi w\). For all junctions, gold wires for the electrical leads are attached using room-temperature-cure silver paste (Dupont, 4922N). The transport measurements are performed with a ${}^{3}$He cryostat down to 300~mK. The cryostat was magnetically shielded with high-permeability material (Hamamatsu Photonics, mu-metal).

\section{Results}
\subsection{Temperature-dependent resistance}
\begin{figure}
	\begin{center}
		\includegraphics[width=8cm]{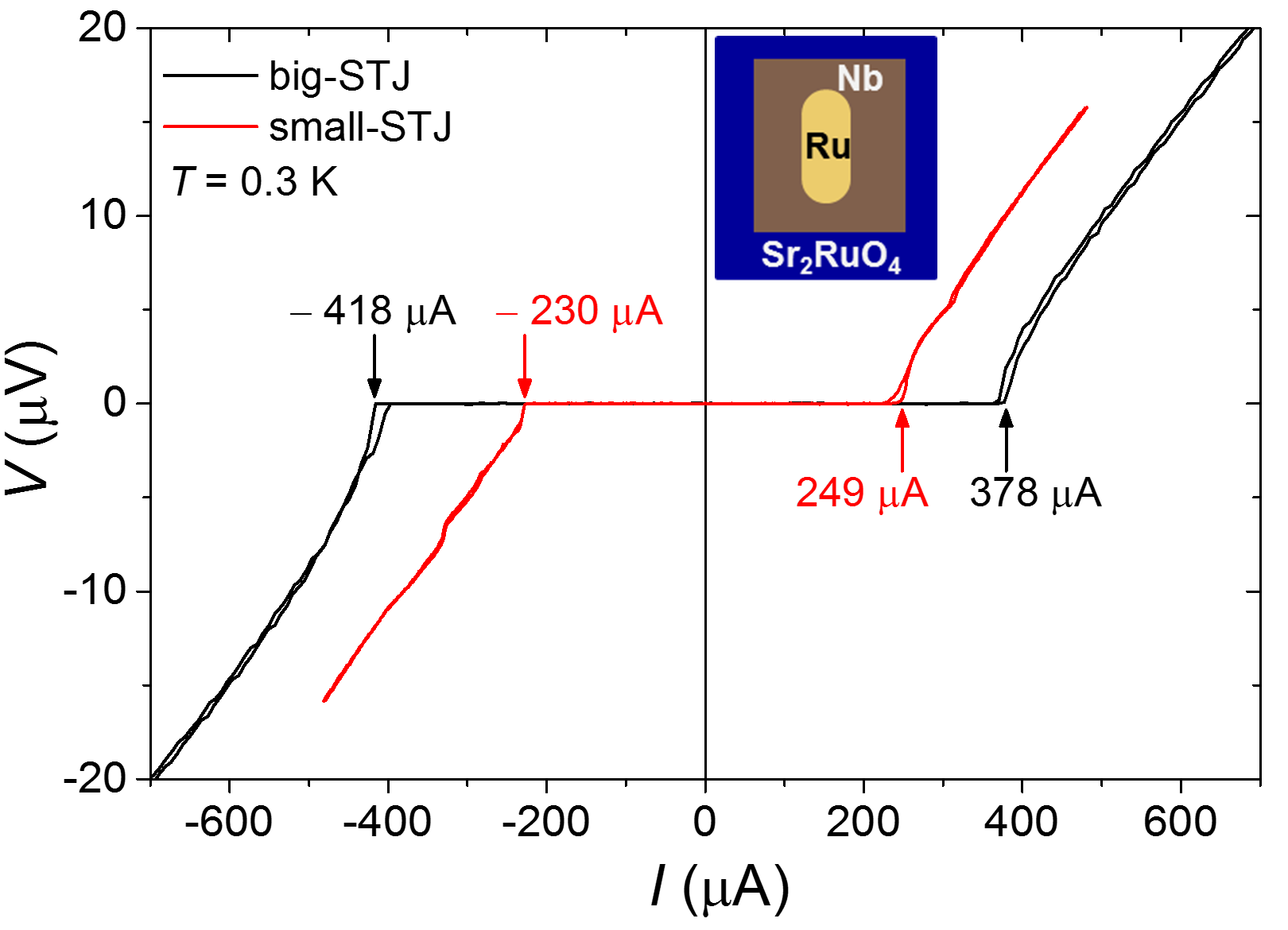}
		\caption{Asymmetric $I$-$V$ curves at 0.3~K of big-STJ (black) and small-STJ (red) in a stable state with small hystereses. Note that the asymmetry is larger for bigger junction. Inset shows the schematic of the junction.}
		\label{IVsSTJ}
	\end{center}
\end{figure}
\begin{figure*}
	\begin{center}
		\includegraphics[width=12cm]{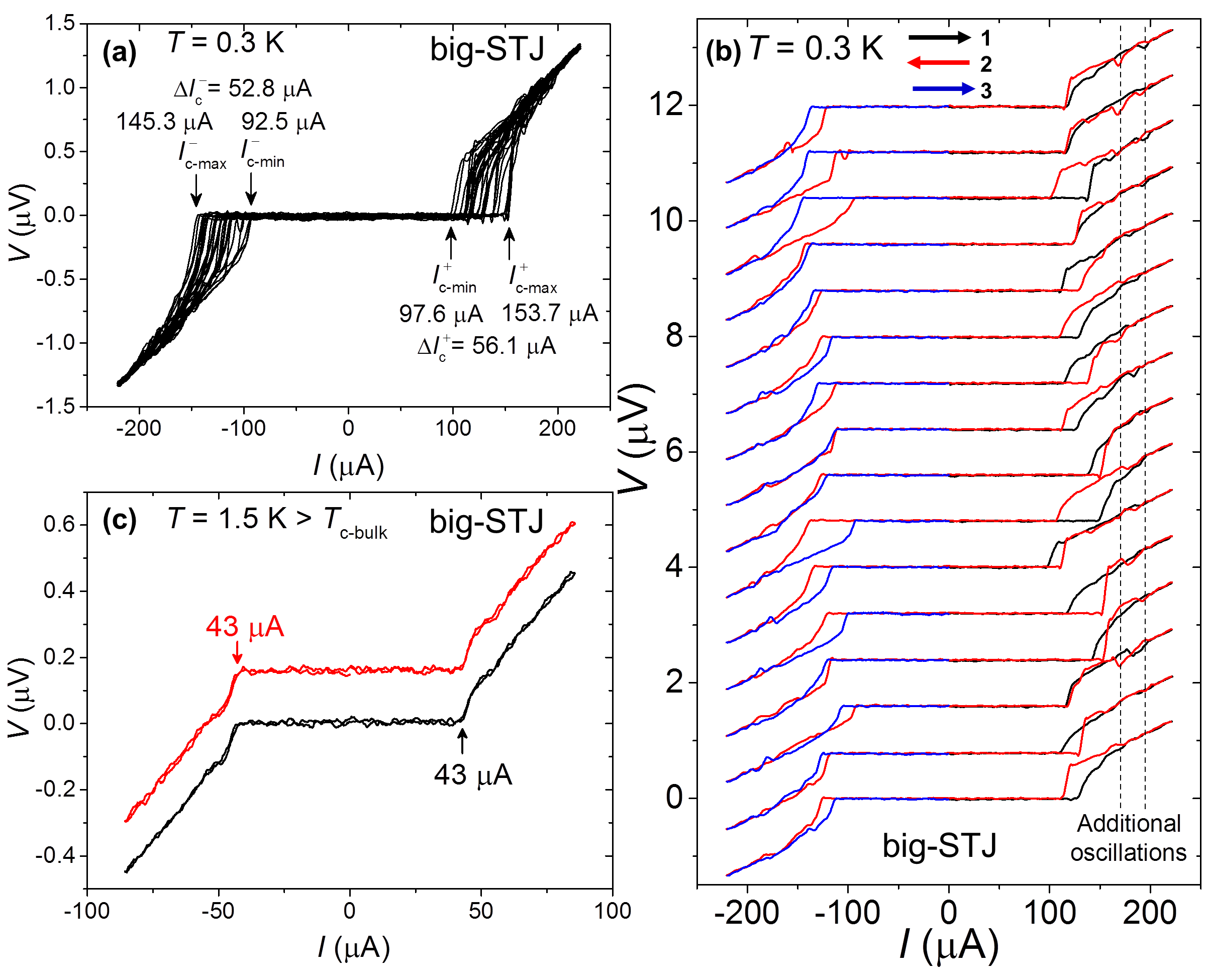}
		\caption{Unstable and stable behavior of 1.5-K and 3-K phases respectively of the big STJ. (a) Sixteen consecutively measured $I$-$V$ curves at 0.3~K, collected in one cooling cycle where the junction is happened to be in an unstable state. Consecutive change in critical current values randomly spread over more than 50 $\mu$A. (b) $I$-$V$ curves shown in (a) with vertical offset illustrating the anomalous hysteretic behavior. Vertical dotted lines indicate the two additional oscillations above the critical current. (c) Two $I$-$V$ curves at 1.5~K measured during two different cooling cycles. The curves are obviously reproducible, symmetric and non-hysteretic.}
		\label{big-STJ}
	\end{center}
\end{figure*}

Figure~\ref{RT}(a) presents the temperature dependent resistance data of a big and a small STJs. Different superconducting transitions are observed. The first sharp transition around 9~K is corresponding to $T_{\textrm{c}}$ of Nb that is not much different than the bulk $T_{\textrm{c}}$ of Nb. It reflects the good quality of the Nb layers. Final transition starts around 2.8~K is initiating from 3-K phase that leads to zero resistance close to 1.8~K (1.6~K) for big-STJ (small-STJ). Normal state resistance defined at 10~K is of ther order of 100-150~m$\Omega$. It indicates the formation of metallic interface between Ru and Nb. Note, zero resistance is achieved significantly above the $T_{\textrm{c-bulk}}$ ($T_{\textrm{c}}$ of 1.5-K phase)~\cite{Anwar2013}.  Figure~\ref{RT}(b) shows the temperature dependent resistance of the equal-DTJ  and ETJ, measured with 30~$\mu$A applied current ($I_{\textrm{a}}$). Multiple transitions are obviously present. For all of these junctions, the first transition occurs at 9~K corresponding to superconductivity in the Nb electrodes. However, the second and broad transition appears at 5.8~K and 6.5~K for equal-DTJ and ETJ, respectively. This second transition arises due to the induction of spin-singlet superconducting correlations into the Ru metal. At lower temperatures, the transition corresponding to the 3-K phase occurs at 1.5~K and 2~K for equal-DTJ and ETJ, with zero resistance transition at 0.9~K and 1~K, respectively. These junctions exhibit the zero resistance $T_{\textrm{c}}$ lower than the ideal $T_{\textrm{c}}$ of the bulk SRO (1.5~K) due to lower critical current density and/or lower Nb/Ru interface transparency.

Figure~\ref{RT}(c) presents the resistance as a function of temperature for an unequal-DTJ. For this junction, we present the resistance data only below 4~K, focusing on the transitions at lower temperatures since the transitions at higher temperatures are similar to those in the other junctions. Interestingly, a sharp transition occurs at 1.75~K for applied current $I_{\textrm{a}} = 5~\mu$A (blue triangles). On further cooling, normal state resistance is recovered at around 1.4~K and jumps to zero resistance at 1.25~K. This anomalous behaviour is suppressed with the increase in $I_{\textrm{a}}$. Note that the normal state resistance is robust again increase in $I_{\textrm{a}}$. These facts indicate that a strong suppression in the critical current only in the vicinity of the 1.5-K phase transition \cite{Nakamura2011} attributable to topological phase competition between the $s$-wave and $p$-wave superconductivity. Such behavior was not observed in other topological junctions studied here. Although such strong suppression was reported in Refs.~\onlinecite{Nakamura2011,Nakamura2012}, it was not observed in other simpler topological junctions studied here.   

\subsection{Current-voltage curves}

Figure \ref{IVsSTJ} shows current-voltage ($I$-$V$) curves at 0.3~K for the big-STJ (black curve) and small-STJ (red curve); schematic of the junction is shown as the inset of Fig.~\ref{IVsSTJ}. Both junctions exhibit asymmetric $I$-$V$ curves with respect to the direction of the current ($I_{\textrm{c+}} \neq |I_{\textrm{c-}}|$). It is observed that $\Delta I_{\textrm{c}} = I_{\textrm{c+}}-|I_{\textrm{c-}}| =-40~\mu$A and $+19~\mu$A for the big-STJ and small-STJ, respectively. The sign and magnitude of the observed asymmetry ($\Delta I_{\textrm{c}}$) varies with cooling cycles. Below $T_{\textrm{c-bulk}}$, the $I$-$V$ curves in most cases are asymmetric, but always become symmetric above $T_{\textrm{c-bulk}}$ (see Fig.~\ref{big-STJ}c). This asymmetric behavior below $T_{\textrm{c-bulk}}$ is consistent with the previous observations on Pb/Ru/SRO junctions, where Pb electrodes were deposited on many Ru inclusions\cite{Nakamura2011,Nakamura2012}. If such asymmetric $I$-$V$ curves arises due to some inhomogeneity or asymmetry at the interfaces of the junction, the asymmetry should not be altered by thermal cycles. On the other hand, some trapped vortices and/or current crowding at the interface can also give rise to asymmetric $I$-$V$s. If it is the case, the asymmetry must present even in the 3-K phase. These are in contrast with the present and previous observations~\cite{Nakamura2011,Nakamura2012,Anwar2013}. Thus, the asymmetry is probably attributable to existence of chiral-DWs: our junctions exhibit different chiral-DW configurations depending on current sweeps and cooling cycles.

\begin{figure}
	\begin{center}
		\includegraphics[width=8cm]{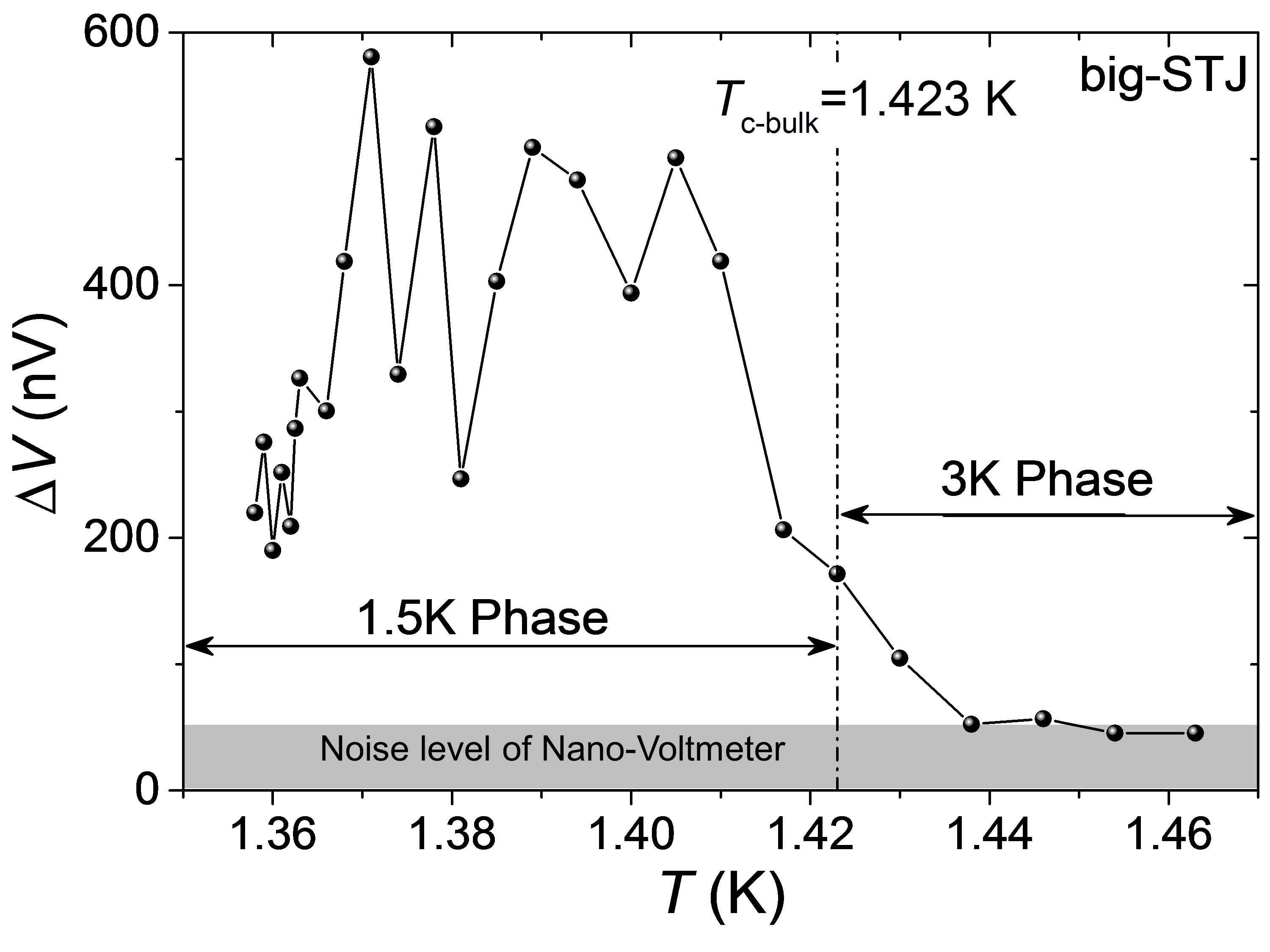}
		\caption{Amplitude of telegraphic-like noise as a function of temperature of the big STJ. It was suppressed almost down to the noise level of the instruments just above 1.42~K, which is exactly {\it T}$_{\textrm{c-bulk}}$ of the Sr$_2$RuO$_4$ crystal.}
		\label{VtT}
	\end{center}
\end{figure}

Interestingly, these junctions do not show the same behavior for every cooling cycle. Figure~\ref{big-STJ}a represents sixteen $I$-$V$ curves of a big-STJ measured consecutively (without disturbing temperature) at 0.3~K during another cooling cycle different from that presented in Fig.~\ref{IVsSTJ}. This time, $I_{\textrm{c}}$ is reduced by more than 50$\%$ and also unstable with the variations more than $\approx$50~$\mu$A. These $I$-$V$ curves also exhibit the unstable and unconventional hysteretic behavior ($I_{\textrm{c}}$ is smaller than the retrapping current $I_{\textrm{r}}$; $I_{\textrm{c}}$<$I_{\textrm{r}}$), see Fig.~\ref{big-STJ}(b). Only three $I$-$V$ curves out of sixteen exhibit normal hysteretic behavior ($I_{\textrm{c}}>I_{\textrm{r}}$). Note that above $I_{\textrm{c}}$ there are also additional oscillations (mainly two, indicated by the dotted black lines in Fig.~\ref{big-STJ}b). This oscillation demonstrates that the dynamic behavior is present above the critical current of the junction. Such an unstable and hysteretic behavior is observed only below 1.42~K (1.5-K phase). Figure~\ref{big-STJ}c shows two $I$-$V$ curves measured at 1.5~K during two different cooling cycles and in both cases the $I$-$V$ curves are rather stable with no asymmetry and hysteresis. 
\begin{figure}
	\begin{center}
		\includegraphics[width=8cm]{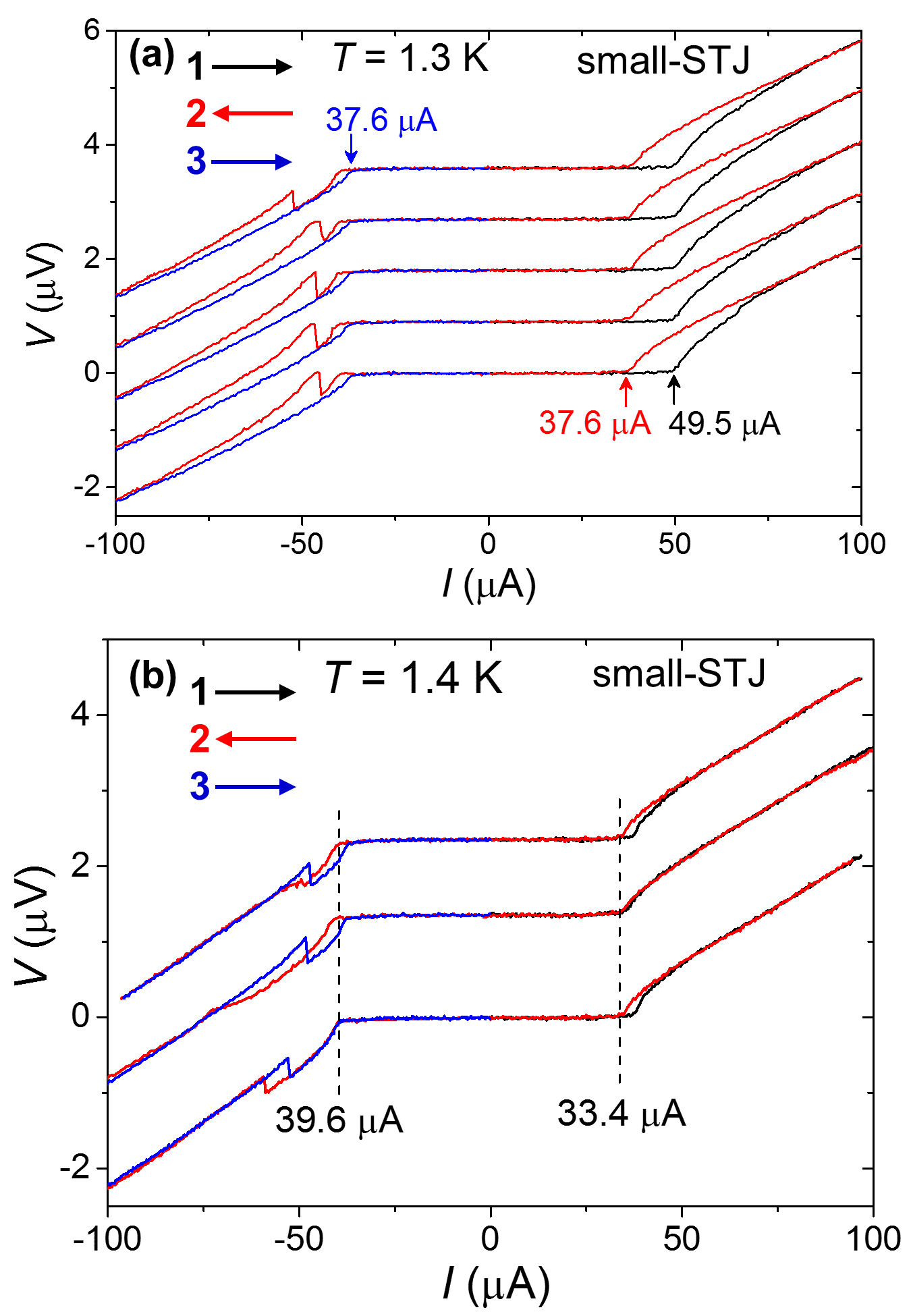}
		\caption{Relatively stable behavior of the small-STJ. (a) Five consecutively measured $I$-$V$ curves at 1.3~K. All curves exhibit stable behavior with normal hysteresis with positive current, but with an anomalous transition appeared during current-increasing for negative current. (b) Consecutively measured three $I$-$V$ curves at 1.4~K. The hysteretic behavior is suppressed but the anomalous transition for negative current still occurs.}
		\label{small-STJ}
	\end{center}
\end{figure}
In order to clarify the temperature evolution of the stability, we investigated the voltage as a function time (telegraphic-like noise) at constant applied current while slowly increasing the temperature from 1.36~K to 1.47~K. The applied current is small enough so that $V$=0 even at 1.47 K. One normally anticipate that $V$ stays zero at lower $T$ because $I_{\textrm{c}}$ increase at lower $T$. Interestingly enough, Fig.~\ref{VtT} shows that dynamic behavior is strongly suppressed at 1.423~K, ${\it T}_{\textrm{c-bulk}}$ of SRO single crystals used in these experiments. Furthermore, noise amplitude is higher close to the transition and decreases at lower temperatures with enhanced $I_{\textrm{c}}$. 
\begin{figure*}
	\begin{center}
		\includegraphics[width=12cm]{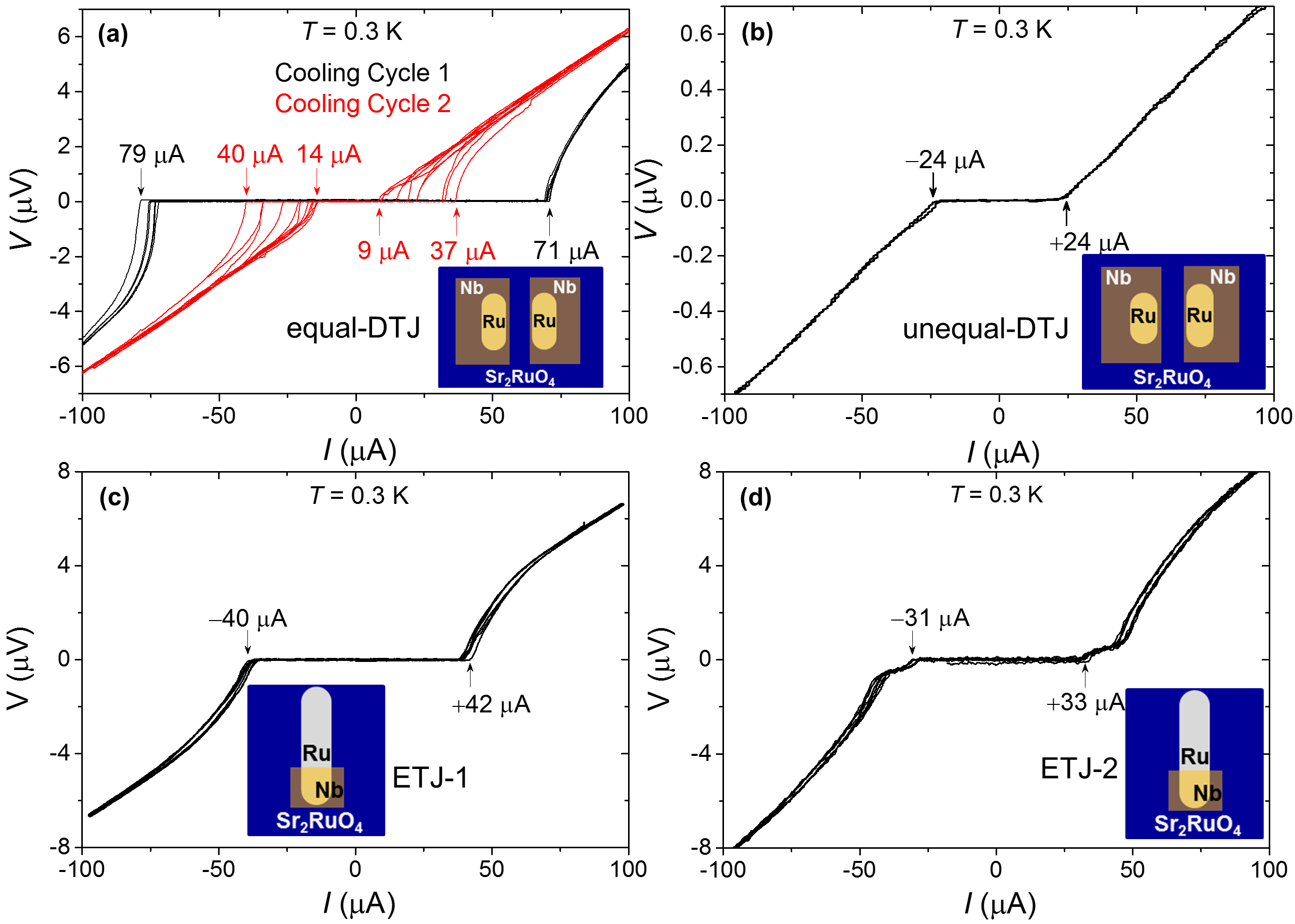}
		\caption{Unstable and stable behavior of DTJs and ETJs. (a) Two different sets of $I$-$V$ curves at 0.3~K measured in two different cooling cycles of an equal DTJ. Black curves with higher $I_{\textrm{c}}$ are rather stable. In contrast, red curves exhibit lower and strongly unstable $I_{\textrm{c}}$ values spread over $\approx$30~$\mu$A. (b) Stable and symmetric $I$-$V$ curves of an unequal-DTJ. Five consecutively measured $I$-$V$ curves of (c) ETJ-1 and (d) ETJ-2. Insets illustrate the top view of corresponding junctions.}
		\label{DTJ}
	\end{center}
\end{figure*}
The small-STJ shows rather stable $I$-$V$ curves at low temperatures in almost all cooling cycles. However, as represented in Fig.~\ref{small-STJ}(a) at 1.3~K close to 3-K phase, the $I$-$V$ curves are persistently and strongly hysteretic with normal behavior ($I_{\textrm{c}}>I_{\textrm{r}}$). Interestingly, a sharp and anomalous transition is observed during current-increasing sweep with negative current. Figure~\ref{small-STJ}b presents three consecutive $I$-$V$ curves at 1.4~K but during the another cooling cycle. This time, hysteresis for positive current is suppressed but anomalous transitions occur for both forward and backward sweeps of negative current. These results suggest that stability of topological junctions has improved at low temperatures (0.3 K) with the reduction of the junction area, as shown in Fig.~\ref{IVsSTJ}; but at higher temperatures (close to the 1.5-K phase transition), they are still relatively unstable.
 
Previously, most of the studies on Pb/Ru/Sr$_2$RuO$_4$ junctions use Pb patches deposited over many Ru-inclusions. In that case, there are many parallel topological junctions with different cross-sectional areas. In contrast, in the present study and in Ref.~\onlinecite{Anwar2013} the devices consist of well-defined single or double topological junctions. Interestingly, in all of these studies result in  normal junction behavior only above $T_{\textrm{c-bulk}}$;
below $T_{\textrm{c-bulk}}$ the junction $I_{\textrm{c}}$ is unstable with asymmetric $I$-$V$s. Nevertheless, only weak signatures of $I_{\textrm{c}}$ suppression is observed in the latter studies with well-defined single or double junctions.

To further understand and control the dynamic behavior of topological junctions, we investigated the DTJs and ETJs. Figure~\ref{DTJ}(a) shows the $I$-$V$ curves of an equal-DTJ (schematic of the junction is given in the inset of Fig.~\ref{DTJ}(a)) measured at 0.3~K for two different cooling cycles. Obviously, strong instability in $I_{\textrm{c}}$ with variation of more than $\approx 30~\mu$A is observed in the junctions. $I$-$V$ characteristics depend on cooling cycle as well, although the size of one Ru inclusion is $6\times2~\mu$m$^2$ (of the order of junction area of small-STJ, which is rather stable). In the stable state (black curves), $I$-$V$ curves are only weakly asymmetric and no hysteretic behavior is observed. Interestingly, the strong stability is achieved by reducing the size of one of the Ru-inclusions down to $4\times2~\mu$m$^2$ (second Ru-inclusion has the same size of $6\times2~\mu$m$^2$). But this junction (unequal-DTJ) shows rather low critical current even at low temperatures (see Fig.~\ref{DTJ}(b)). This low critical current is also seen in the resistance behavior; strong suppression of $T_{\textrm{c}}$ even with 20~$\mu$A (Fig.~\ref{RT}(c)). These results suggest that the instability in equal-DTJ arises due to chiral-DW interaction and current distribution around the all round parts of both Ru-inclusions 

Figures~\ref{DTJ}(c) and (d) present $I$-$V$ curves (five consecutive loops) of two different ETJs but of the same junction area ($3\times2~\mu$m$^2$). It is clear that for both of these junction the $I$-$V$ characteristic curves are persistently stable. There still is a very weak normal hysteretic behavior and asymmetry. Nevertheless, these facts, suggest that ETJs exhibit higher stability.

\section{Discussions}

Before starting our discussion, we briefly summarize our main experimental results related to the stability of our topological junctions fabricated in various configurations. The big-STJ in most cases exhibits unstable behavior with $\approx 60\%$ reduction in critical current compaired with that in a stable state. In the unstable state, the $I$-$V$s exhibit unconventional hysteretic behavior ($I_{\textrm{c}}<I_{\textrm{r}}$) but only below $T_{{\textrm{c-bulk}}}$. Above $T_{\textrm{c-bulk}}$, the $I$-$V$s are rather stable, symmetric and non-hysteretic. The stability can be achieved with the reduction of the junction area down to $6\times2~\mu$m$^2$ the small-STJ. Stability of topological junctions can be controlled also by changing the configuration of the junctions, e.g. equal-DTJ are unstable relative to unequal-DTJ but ETJ is completely stable. Note that instability depends on cooling cycle as well.

\begin{figure}
	\begin{center}
		\includegraphics[width=6cm]{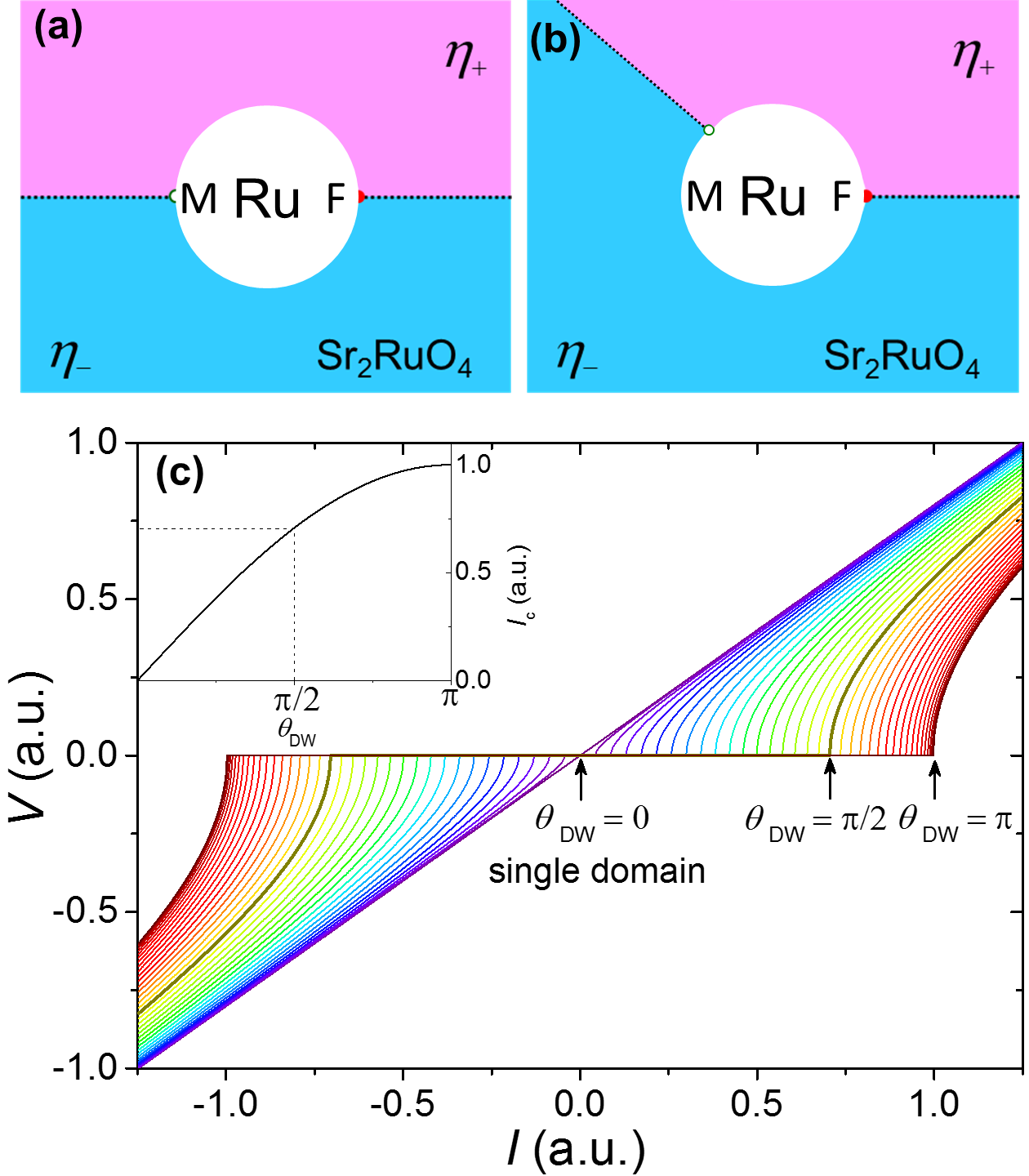}
		\caption{Critical current based on a model with chiral-domain wall (chiral-DW) motion for a STJ. (a) Chiral-DW configuration for SRO-Ru system: the circular white part is the Ru-inclusion surrounded by two chiral domains with different chirality separated by two chiral-DWs; one chiral-DW 'F' at $\theta = 0$ is assumed to be pinned and the other chiral-DW 'M' at $\theta_{\rm DW}$. Maximum sustainable current can be achieved when pinning M at $\theta_{{\textrm{DW}}} = \pi$. (b) By pinning M at $\theta_{\rm DW} = \pi/2$, the critical current is reduced by 30\%. (c) $I$-$V$ curves calculated varying $\theta_{{\textrm{DW}}}$ from $0$ to $\pi$. Inset shows the critical current as a function of $\theta_{{\textrm{DW}}}$.}
		\label{cal}
	\end{center}
\end{figure}

Our results reveal that the stability is dependent on junction size, the bigger the more unstable as summarized in Table 1. Furthermore, the unstable behavior emerges only below $T_{{\textrm{c-bulk}}}$. Such a dynamic behaviour is consistently explained by the motion of chiral-DWs of the SRO spin-triplet superconductor. Below, we perform calculations using a simple model based on two chiral-DWs to simulate our results. We consider two chiral-DWs separating two chiral domains with opposite chirality and intersecting a circular Ru-inclusion as illustrated in Fig.~\ref{cal}(a) and \ref{cal}(b). For simplicity, we fixed one chiral-DW 'F' at $\theta = 0$, while the other chiral-DW 'M' at some angle $\theta=\theta_{{\textrm{DW}}}$ is free to move between stable and metastable states. For such a chiral-DW configuration, $I_{\textrm{c}}$ can be calculated using the following relation~\cite{Anwar2013}, 

\begin{equation}
\begin{split}
I_\textrm{c} =\textrm{max}[\frac{I_{\textrm{co}}}{2\pi}{\int_{0}^{\theta_{\textrm{DW}}}d\theta \textrm{sin} \varphi_{+}(\theta; \theta_{{\textrm{DW}}}, \delta\varphi)}\\ { +\int_{\theta_{{\textrm{DW}}}}^{2\pi}d\theta \textrm{sin} \varphi_{-}(\theta; \theta_{{\textrm{DW}}}, \delta\varphi)}]
\end{split}
\label{eq}
\end{equation}

\noindent where, $\delta\varphi$ is the phase difference at $\theta=0$ between an $s$-wave spin-singlet superconductivity (induced into Ru from Nb electrode) and $p$-wave spin-triplet superconductivity SRO. For the single-valuedness and symmetry of the order parameter, we consider the phase difference across a chiral-DW \(\alpha = \pi - \theta_{\textrm{DW}}\) and \(\alpha_{{\textrm{M}}} = -\alpha_{{\textrm{F}}}\) (see Figs.~\ref{cal}(a)) and ~\ref{cal}(b) and $\delta\varphi$ is varied so that $I_\textrm{c}$ becomes maximal. Calculated $I$-$V$ curves using the relation for an overdamped junction and $I_{\textrm{c}}$ at various $\theta_{\textrm{DW}}$ varying from $0 - \pi$ are given in Fig.~\ref{cal}(c). The sinusoidal variations of $I_\textrm{c}$ as function of $\theta_{\textrm{DW}}$ is shown in the inset of Fig.~\ref{cal}(c). The maximum $I_{\textrm{c}}$ is found for M at position $\theta_{{\textrm{DW}}} = \pi$ and $\delta\varphi = \alpha = 0$ as shown in Fig.~\ref{cal}(a). The $I_\textrm{c}$ reduces down by 30\% for $\theta_{{\textrm{DW}}} = 0.5\pi$ ($\delta\varphi = 0.25\pi$ and $\alpha = 0.5\pi$); this configuration is shown schematically in Fig.~\ref{cal}(b). The 60\% suppression in the $I_\textrm{c}$ is achieved at $\theta_{{\textrm{DW}}} = 0.3\pi$ ($\delta\varphi = 0.35\pi$ and $\alpha = 0.7\pi$); such a reduction corresponds to the small $I_\textrm{c}$ of the big-STJ in the unstable state.

\begin{figure}
	\begin{center}
		\includegraphics[width=6cm]{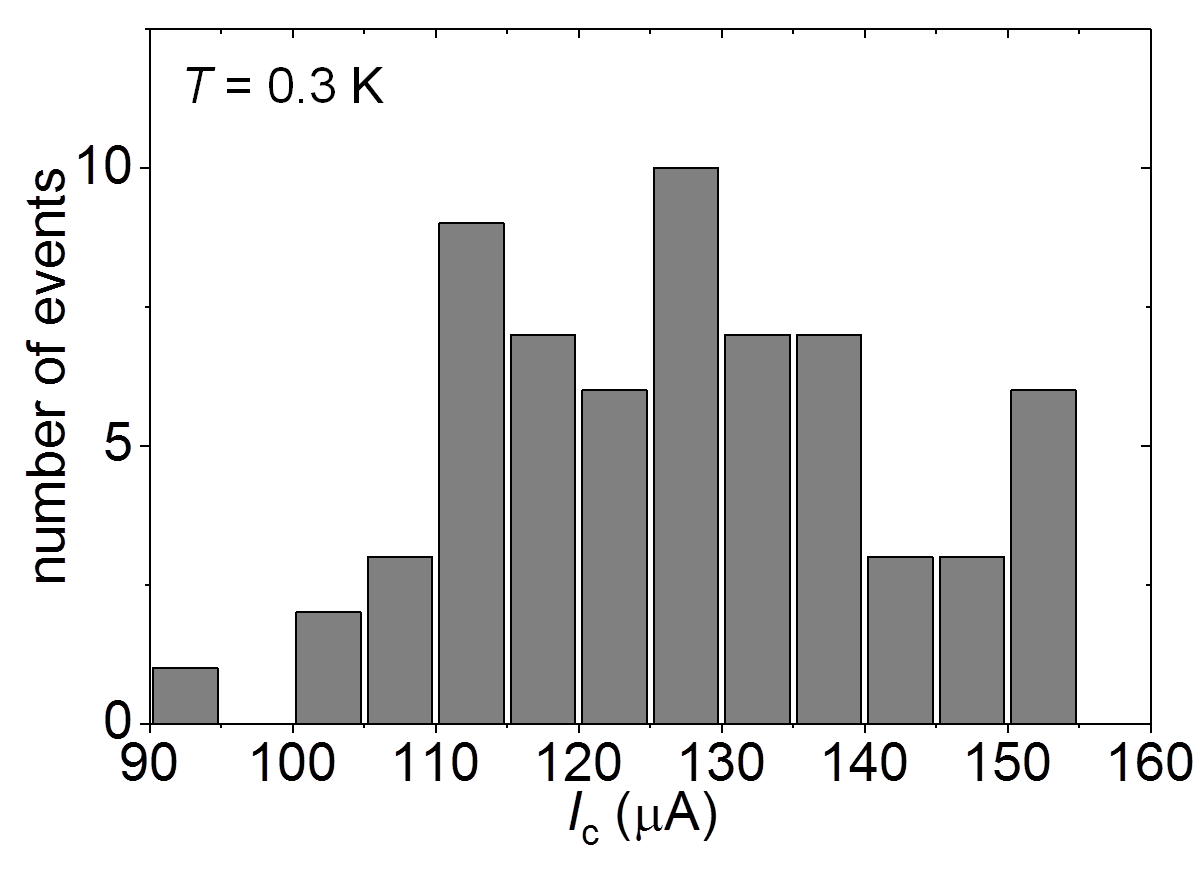}
		\caption{Histogram of critical current measured from hysteretic $I$-$V$s of the big-STJ given in Fig. \ref{big-STJ}b.}
		\label{histo}
	\end{center}
\end{figure}

\begin{table*}[ht]
	\centering
	\caption{Properties of junctions at 0.3 K (sorted by the $L_{\textrm{s}}/L_{\textrm{c}}$ ratio)}
	\begin{tabular}{|c|c|c|c|c|c|c|c|c|c|c|}
		\hline
		
		& \vtop{\hbox{\strut Ru size}\hbox{\strut ($\mu$m$^2$)}} &\vtop{\hbox{\strut $L_{\textrm{s}}$}\hbox{\strut ($\mu$m)}} &\vtop{\hbox{\strut $L_{\textrm{c}}$}\hbox{\strut ($\mu$m)}} & $L_{\textrm{s}}$/$L_{\textrm{c}}$&\vtop{\hbox{\strut stable/unstable}\hbox{\strut switching}} & state & asymmetry& \vtop{\hbox{\strut $I$-$V$ ordinary}\hbox{\strut hysteresis}} & \vtop{\hbox{\strut $I$-$V$ unusual}\hbox{\strut hysteresis}} & Instability\\
		\hline
		
		{\multirow{2}{*}{big-STJ}}& {\multirow{2}{*}{2$\times$20}} &{\multirow{2}{*}{36}}&{\multirow{2}{*}{6.3}}&{\multirow{2}{*}{ 5.7 }}&{\multirow{2}{*}{yes}} & unstable & yes & yes & yes &yes\\
		
		&  & & & &  &stable &yes& small & no &no\\  
		\hline 
		{\multirow{2}{*}{equal-DTJ}}  & {\multirow{2}{*}{2$\times$6, 2$\times$6 }} &{\multirow{2}{*}{16}}&{\multirow{2}{*}{12.6}} & {\multirow{2}{*}{1.3}}&{\multirow{2}{*}{yes}} &unstable&yes& yes & yes & yes\\
		
		& & & & & & stable & yes & small & no & no\\  
		\hline  
		small-STJ   & 2$\times$5 &6&6.3& 1.0&no & stable &yes& yes & no &small\\
		\hline  
		unequal-DTJ    &2$\times$6, 2$\times$4 &12&12.6& 1.0&no & stable & no & no & no & no\\
		\hline
		ETJ-1& 2$\times$3 &2&6.3& 0.3&no & stable &small& small & no &no\\
		\hline
		ETJ-2      & 2$\times$3 &2&6.3&0.3&no & stable &small& small & no &no\\ 
		\hline
	\end{tabular}
	\label{table}
\end{table*}

Realistically, the Ru-inclusions embedded in SRO crystals used in this work are in a rectangular shape with two circular edges and straight parts (see Fig.~\ref{devices}(j)). The edges must have higher crystal mismatch between SRO and Ru due to larger curvature. Such mismatch results in higher pinning potential providing suitable pinning sites to pin the chiral-DWs. In this scenario, the smooth straight parts of Ru-inclusions have comparatively lower pinning potential. Therefore, the most stable configuration with maximum $I_\textrm{c}$ is obtained by pinning two chiral-DWs at the tips of opposite circular edges of a Ru-inclusion. Our simple model calculations also predict that the stability with maximum $I_\textrm{c}$ is obtained when the chiral-DW M is pinned at $\theta_{{\textrm{DW}}} = \pi$. The stability of the junction is disturbed with the increase in the temperature and external magnetic field scans as well~\cite{Kindwingira2006,Anwar2013}. But high enough $I_\textrm{c}$ ($\approx 400~\mu$A) can itself be a dragging-force to move the chiral-DWs from a stable state to neighbouring metastable states that results in reduction and fluctuations in $I_\textrm{c}$. Nevertheless, it would be difficult to eliminate the chiral-DW since the chirality is degeneracy of the angular momentum direction of Copper pairs and does not couple with electric current strongly. Such dynamic effects can be the origin of the unconventional hysteretic behaviour ($I_{\textrm{c}} > I_{\textrm{r}}$) and variations in $I > I_\textrm{c}$ by switching a chiral-DW to another close by metastable state. That means chiral-DW motion can be initiated in criticality region (between zero voltage transition and straight ohmic behavior). This is also supported by the observations of oscillations just above $I_\textrm{c}$ in $I$-$V$ curves for STJs in unstable state (see Fig.~\ref{big-STJ}(b)).  

The stability in the Nb/Ru/SRO junctions is expected to be achieved by reducing the size of the junction down to the size of the chiral domain. The small-SJT junction with the junction area of $2 \times 5 \mu$m$^2$ is rather stable. It suggests that the size of a chiral domain is of the order of $5~\mu$m~\cite{Saitoh2015}. Interestingly, small-STJ shows the sharp switching in $I$-$V$ curves at 1.3~K only for negative current (Fig. 6). Such an unusual $I$-$V$ characteristics behavior is already reported by Kambara $et~al.$,~\cite{Kambara2008,Kambara2010} in Ru/SRO junctions. Those results were attributed to the current-induced motion of chiral-DWs in 3-K phase.

We summarise the device configuration parameters and results of stability in Table~\ref{table}. It revels that the devices having higher $L_{\textrm{s}}/L_{\textrm{c}}$ ratio are more unstable. It suggests that the stability can be achieved with increasing the length of the curved length $L_{\textrm{c}}$ of the Ru-inclusion.  On the other hand, Eq~(\ref{eq}) shows that the direction ($\theta$ is defined as normal to the Ru/SRO interface) is constant for a flat part, and strongly varying for curved part of the Ru-inclusion. In addition, the pinning potential should be lower for the flat part and higher for the curved part due to larger lattice mismatch along the curvature. These considerations suggest that the stability of Nb/Ru/SRO topological junctions can be controlled by correctly selecting the part of the Ru-inclusion to place the Nb electrode. These are our results, DTJs and ETJs.

Our systematic investigations suggest that the switching in the $I_{\textrm{c}}$ due to chiral-DWs motion between neighbouring metastable states, can vary $I_{\textrm{c}}$ in discrete current values rather than continuously. Figure~\ref{histo} shows the histogram of $I_{\textrm{c}}$ variations in big-STJ. Note that we included both $+I_{\textrm{c}}$ and $-I_{\textrm{c}}$ obtained from the data presented in Fig.~\ref{big-STJ}(a and b). In an unstable state $I_{\textrm{c}}$ is mainly distributed between $110~\mu$A and $140~\mu$A. Furthermore, the $I_{\textrm{c}}$ variations are not completely continuous but to claim the discreteness more data may be needed. That will be future of our study; quantum variations in $I_{\textrm{c}}$ due to chiral-DW motion. 

Recently, Etter {\it et al.}~\cite{Etter2014}, proposed that transport properties of topological junctions based on the eutectic SRO-Ru system depend strongly on phase winding. They found that in the 3-K phase the phase winding is zero and junction exhibits ordinary or unfrustrated behaviour. However, the phase winding is non-zero ($\pm 1$) in the 1.5-K phase and junction is frustrated due to the phase mismatch between Ru and SRO. This proposed behavior is similar to our experimental observation. Specifically, this theory suggests stability will be enhanced with increase in the Ru size. In contrast, we observed that larger Ru size shows more unstable behavior.

Before closing discussion, we would like to comment on the effect of possible vortex trapping. Some vortices can be trapped in the junction even under a very small residual field during cooling down below $T_{\textrm{c}}$. The dynamics of such trapped vortices can cause anisotropic $I$-$V$ curves due to the vortex flow from the Ru/SRO interface to the bulk SRO, or vice versa~\cite{Lee2002}. However, it is difficult to explain the anomalous hysteresis~\cite{Yu1993} (see Fig. 4b). Furthermore, the vortex flow, if playing a crucial role, should be observable below the onset temperature of the 3-K phase or at least below the temperature where zero resistance emerges (1.8~K in the case of STJs). In contrast, the observed asymmetric $I$-$V$ curves with anomalous hysteresis and telegraphic-like noise are observed only when the bulk superconductivity in SRO (see Figs. 4 and 5) sets in. In addition,  we again emphasize that we cooled down the junctions rather slowly in a zero-field environment prepared by magnetic shielding in order to suppress any significant effect of trapped vortices.

\section{Conclusion}

We investigated Nb/Ru/SRO topological junctions fabricated in various configurations, sizes and shapes of the Ru inclusions. Such superconducting junctions exhibit contrasting instabilities in the critical currents between the $s$-wave proximitized Ru and SRO. Junctions with relatively large Ru inclusions are rather unstable and exhibit a large noise, asymmetry in current reversal, and hysteresis in current-sweep loops and on cooling cycles. In contrast, junctions with a size smaller than about 5~$\mu$m exhibit ordinary stable $I$-$V$ characteristics. A striking disappearance of the noisy character is observed when the junction is slowly heated across the bulk $T_{\textrm{c-bulk}}$ into the 3-K interfacial superconductivity region.

All these results are coherently explained if the bulk superconducting phase (the 1.5-K phase) of SRO has multi-component order parameter resulting in superconducting domain structure. This is consistent with the chiral $p$-wave superconductivity of the bulk SRO and the non-chiral superconductivity of the 3-K phase. We systematically investigated various junctions but still more statistics is needed to have a firm conclusion. Our work will stimulate research work to explore the physics of topological superconducting junctions and functionality of their dynamic behavior.

\section{Acknowledgment}
We thank S. Kashiwaya, and Rina Takashima for fruitful discussions. This work is supported by the Topological Quantum Phenomena (Nos. JP22103002 and JP25103721) and Topological Materials Science (JSPS KAKENHI JP15H05852) on Innovative Areas from the Japan Society for the Promotion of Science (JSPS). MSA is supported as an International Research Fellow of the JSPS.

\end{document}